\documentclass[%
 reprint,
 amsmath,amssymb,
 aps,
]{revtex4-2}
\usepackage[ngerman,english]{babel}
\usepackage{graphicx}
\usepackage{dcolumn}
\usepackage{comment}
\usepackage{bm}
\usepackage{lineno}

\newcommand{\beginsupplement}{%
        \setcounter{table}{0}
        \renewcommand{\thetable}{A\arabic{table}}%
        \setcounter{figure}{0}
        \renewcommand{\thefigure}{A\arabic{figure}}%
        \setcounter{equation}{0}
        \renewcommand{\theequation}{A.\arabic{equation}}

     }
     
\begin{document}
\selectlanguage{English}

\title{Time-Spatial Mode Selective Quantum Frequency Converter}

\author{Santosh Kumar}
\email{skumar5@stevens.edu}
\author{He Zhang}%
\author{Prajnesh Kumar}%
\author{Malvika Garikapati}
\author{Yong Meng Sua}%
 \author{Yu-Ping Huang}%
\email{yhuang5@stevens.edu}

\affiliation{Department of Physics, Stevens Institute of Technology, Hoboken, NJ, 07030, USA}%
\affiliation{Center for Quantum Science and Engineering, Stevens Institute of Technology, Hoboken, NJ, 07030, USA}

\date{\today}

\begin{abstract}
We experimentally demonstrate a mode-selective quantum frequency converter over a compound spatio-temporal Hilbert space. We show that our method can achieve high-extinction for high-dimensional quantum state tomography by selectively upconverting the signal modes with a modulated and delayed pump. By preparing the pump in optimized modes through adaptive feedback control, selective frequency conversion is demonstrated with up to 30 dB extinction. The simultaneous operations over high-dimensional degrees of freedom in both spatial and temporal domains can serve as a viable resource for photon-efficient quantum communications and computation. 
\end{abstract}



\maketitle

\section{Introduction}
High-dimensional (HD) quantum information carriers have been studied with superconducting circuits \cite{Liu2017}, atomic ensembles \cite{Ding2016,Ding_PhysRevA_HD_atoms,PhysRevA.101.032312}, photons \cite{Sit:17,Singh:19}, and so on. They can increase the information capacity, improve the noise resistance, and make the quantum cryptographic schemes more difficult to hack despite errors \cite{Otte:20,RevModPhys.74.145}. 
In photonic systems, single photons can carry quantum information in a HD Hilbert space subtending polarization, spatial, temporal, frequency, and path degrees of freedom (DOF), with broad applications in long-distance quantum communication, quantum key distribution, quantum gate operations, and quantum teleportation \cite{Xie2015,Erhard2018,Pan_PRL19,Brandt:20,PhysRevLett.125.230501,HDEntanglement_PRL2020,PhysRevLett.125.230501}. 
 
Meanwhile, in modern quantum information architectures, quantum frequency conversion (QFC) has become an essential element with numerous applications in quantum communications and quantum information processing \cite{huang_mode-resolved_2013, manurkar_multidimensional_2016}. It allows us to convert the frequency of the photons without disturbing their quantum properties \cite{Kumar:90}. Through QFC, quantum information can be transmitted through optical fiber links using telecom wavelength at low loss while interfaced with quantum atomic memories \cite{Kaiser:19,PhysRevLett.124.010510,Bock2018}. Many previously studied nonlinear optics based QFCs have been demonstrated in two-dimensional Hilbert space for qubit teleportation and communications \cite{VanDevender:07,Farias2015,Lukens:17,Lu:18,Kagalwala2017}. Recently, this study has been extended to a HD Hilbert space using spatial, frequency, or temporal DOF due to its potential in long distance quantum communications \cite{Fickler2014,Ram_PRApp2019,Xavier2020}. For example, HD QFC was demonstrated for orbital angular momentum (OAM) qudits via sum frequency generation (SFG) with a flat-top Gaussian pump beam \cite{PhysRevA.101.012339}. The HD quantum information of the single photons were shown to be transferred from the polarization DOF to the orbital angular momentum DOF and orbital angular momentum to temporal DOF or vice versa \cite{Nagali:09, Karimi:12, Maring:18}. 



In view of the two vivid directions of pursuit, in this paper, we propose and experimentally demonstrate a HD mode-selective QFC in a compound spatio-temporal Hilbert space. We show that the mutually unbiased basis (MUB) sets in a HD Hilbert space can be selectively up-converted according to their spatial and temporal modes. In contrast to previous demonstrations \cite{PhysRevA.101.012339}, we use an optimized pump beam which is the superposition of many higher-order orthogonal modes. The optimization is performed by using an adaptive feedback technique \cite{Santosh19}. Our HD mode-selective QFC provides an ingenious and effective solution for quantum information processing using single photons in HD Hilbert spaces, with potential applications in quantum communication and quantum networks \cite{Kimble2008,Rempe_RMP15,Liueaay20}. In future, it can be extended to hyper-entangled photon generation for promising quantum applications \cite{Allgaier2017,PhysRevA_Xiang,PhysRevLett_Chen,Tang2020}.

%

\section{Model} 
We consider a signal and a pump beam in the spatio-temporal domain as
\begin{eqnarray}
    \Psi_{i\equiv p,s}(x,y,z,t)=A_{i} E^r_{i}(x,y,z) E^t_{i}(t) \exp[-i(k_{i} z- \omega_{i} t)], ~~~~~
\end{eqnarray}
where $A_{i\equiv p,s}$ is the input electric-field amplitude, and $E^r_{i}(x,y,z)$ and $E^t_{i}(t)$ are the electric fields in the spatial and temporal domain, respectively. Here, we consider that the full electric-field is the product of the spatial and temporal DOF, i.e. we ignore the inter-coupling of the space and time DOF \cite{hancock_free-space_2019,chong_generation_2020}. Under the slowly-varying-envelope approximation with negligible group velocity dispersion, the coupled wave differential equations can be written as: \begin{eqnarray}
2ik_{s}(\partial_z + \beta'_s \partial_t) \Psi_{s}+(\partial_{x}^{2}+\partial_{y}^{2})\Psi_{s}=-2\frac{\omega_{s}^{2}}{c^{2}}\chi \Psi_{p}^{*}\Psi_{f}e^{i\triangle k z},~~~~~
\label{eqES}\\
2ik_{p}(\partial_z+ \beta'_p \partial_t)\Psi_{p}+(\partial_{x}^{2}+\partial_{y}^{2})\Psi_{p}=-2\frac{\omega_{p}^{2}}{c^{2}}\chi \Psi_{s}^{*}\Psi_{f}e^{i\triangle k z},~~~~~
\label{eqEP}\\ 
2ik_{f}(\partial_z+ \beta'_f \partial_t)\Psi_{f}+(\partial_{x}^{2}+\partial_{y}^{2})\Psi_{f}=-2\frac{\omega_{f}^{2}}{c^{2}}\chi \Psi_{p}\Psi_{s}e^{-i\triangle k z}, ~~~~\label{eqEF}
\end{eqnarray}
where $\beta'_{i=s,p,f}$ and $\omega_{i}$ are the inverse group velocities, and frequencies of signal, pump and SF light, respectively.  
$k_{s}={n_{s}\omega_{s}}/{c}$ , $k_{p}={n_{p}\omega_{p}}/{c}$ and $k_{f}={n_{f}\omega_{f}}/{c}$
are their wave numbers. $\chi$ is the second-order nonlinear susceptibility. $\Delta k=k_{s}+k_{p}-k_{f}-2\pi/\Lambda$ is the momentum mismatching condition, where $\Lambda$ is the poling period of the nonlinear crystal. Equations (\ref{eqES})-(\ref{eqEF}) can be solved numerically using the split-step Fourier method with adaptive step-size \cite{QPMS2017,Santosh19}.

\begin{figure}[ht]
    \centering
    \includegraphics[width=\linewidth]{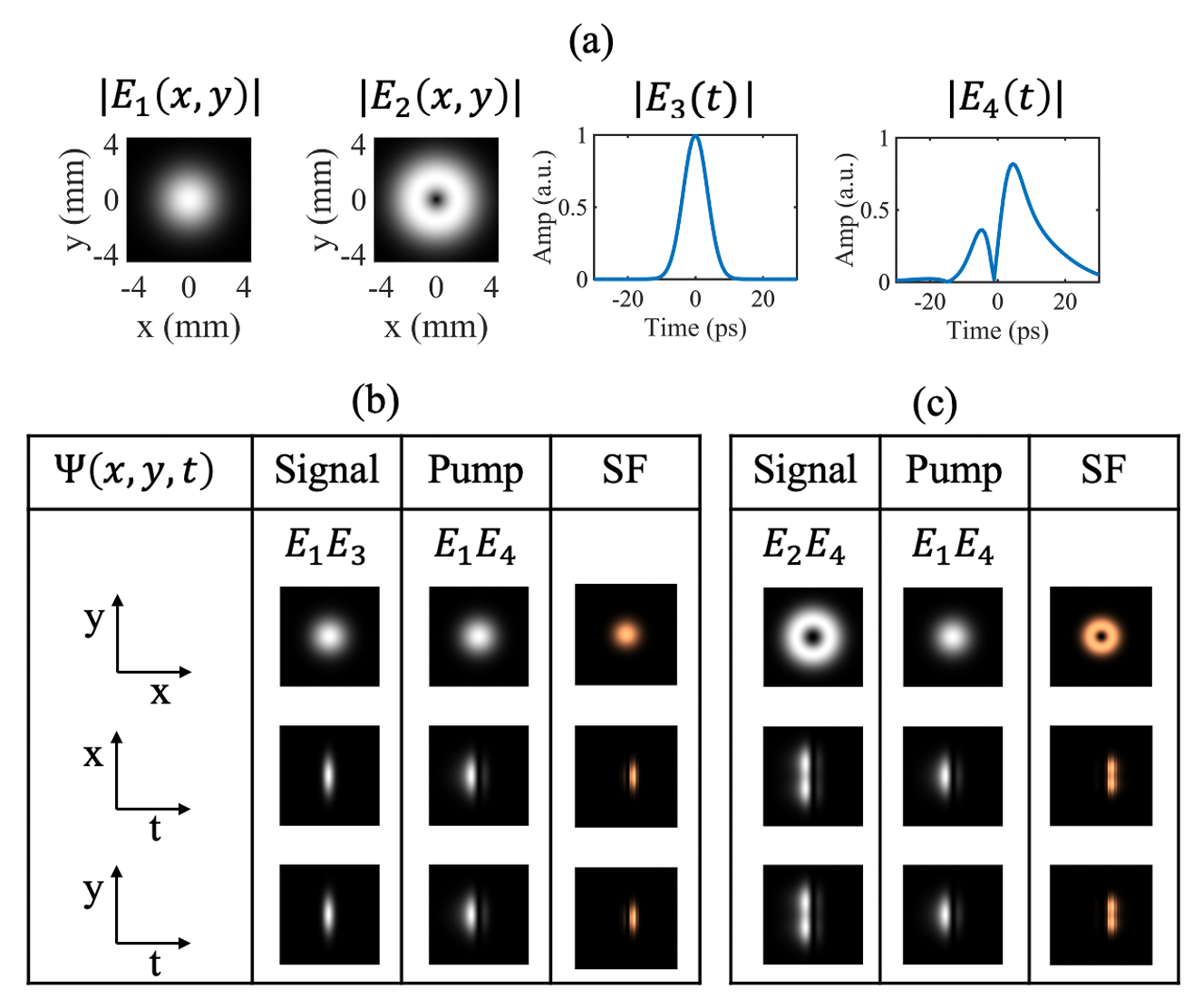}
    \caption{Generated spatio-temporal SF modes. (a) shows the electric field amplitude in the spatial and temporal domain for composing different signal and pump modes. (b) SF output for signal mode $\Psi_{s}=E_1 E_3$ with pump mode $\Psi_{p}=E_1 E_4$. (c) SF output  for signal mode $\Psi_{s}=E_2 E_4$ with pump mode $\Psi_{p}=E_1 E_4$.
     }
    \label{fig:Simulatedtimespatial}
\end{figure}

Figure \ref{fig:Simulatedtimespatial} shows an example of the simulated SF output for two different signal modes in the combined spatial and temporal domain. Both signals are converted by the same spatio-temporal pump field. Figure \ref{fig:Simulatedtimespatial}(a) plots the electric field amplitudes in the spatial and temporal domain. These amplitudes are used to create the spatio-temporal signal and pump fields. Figure \ref{fig:Simulatedtimespatial}(b) and (c) show the SF output for the signal field $\Psi_{s1}=E_1(x,y) E_3(t)$ and $\Psi_{s2}=E_2(x,y) E_4(t)$ with pump field $\Psi_{p}=E_1(x,y) E_4(t)$, respectively. $E_1(x,y)=e^{-\frac{x^2+y^2}{W_0^2}}$ and $E_3(t)=e^{-\frac{(t-t_0)^2}{(\tau_0)^2}}$ are the Gaussian electric field amplitude in the spatial and temporal domain with widths $W_0$ and $\tau_0$, respectively. $E_2(x,y)$ is a Laguerre-Gaussian mode with angular mode number $l = 1$ and radial mode number $p=0$ (see Appendix A). Each plot shows the distribution of the modes in the spatial and temporal domain.

Next, we show that the optimized pump field can selectively upconvert a desired signal mode over other overlapping modes. In general, the optimized pump is considered as a superposition of the higher-order modes. It can be written as
\begin{eqnarray} 
    P_{opt}= [C_{11} LG^0_0 + C_{01} LG^0_1 + ... +C_{pl} LG^p_l(x,y)] \times \nonumber\\ (\tau_1 \Phi_1(t) + \tau_2 \Phi_2(t) + ... + \tau_j \Phi_j(t)), ~~~
\end{eqnarray}
where $LG^p_l$ are the Laguerre-Gaussian modes, $\Phi_j$ are the Schmidt decomposed temporal modes, $C_{pl}$ and $\tau_j$ are the normalized coefficients of the spatial and temporal modes, respectively.

For a given optimized pump, to select a $i^{th}$ signal mode among other overlapping modes. We first define the normalized SF photon, $\bar{N_i}$, of the $i^{th}$ signal as:
\begin{eqnarray} 
   \bar{N_i}=10 log_{10}\left( \frac{N_i}{\sum^n_{i=1} N_i}\right), ~~~
   \label{eq:norm}
\end{eqnarray}
where $N_i$ is the measured SF photons of the $i^{th}$ signal mode and $n$ is the total number of signal modes. Note that, the input incident photons for each signal mode are equal. The selectivity, $\eta$, between desired signal mode and other overlapping modes can be defined as \begin{equation}
\eta=\bar{ N}_i-\bar{N}_{j\neq i},
\label{eq:selectivity}
\end{equation}
where $\bar{N}_{i}$ and $\bar{N}_{j\neq i}$ are the normalized SF photons of the desired signal and the other overlapping signal modes, respectively. 

\begin{figure}[ht]
    \centering
    \includegraphics[width=\linewidth]{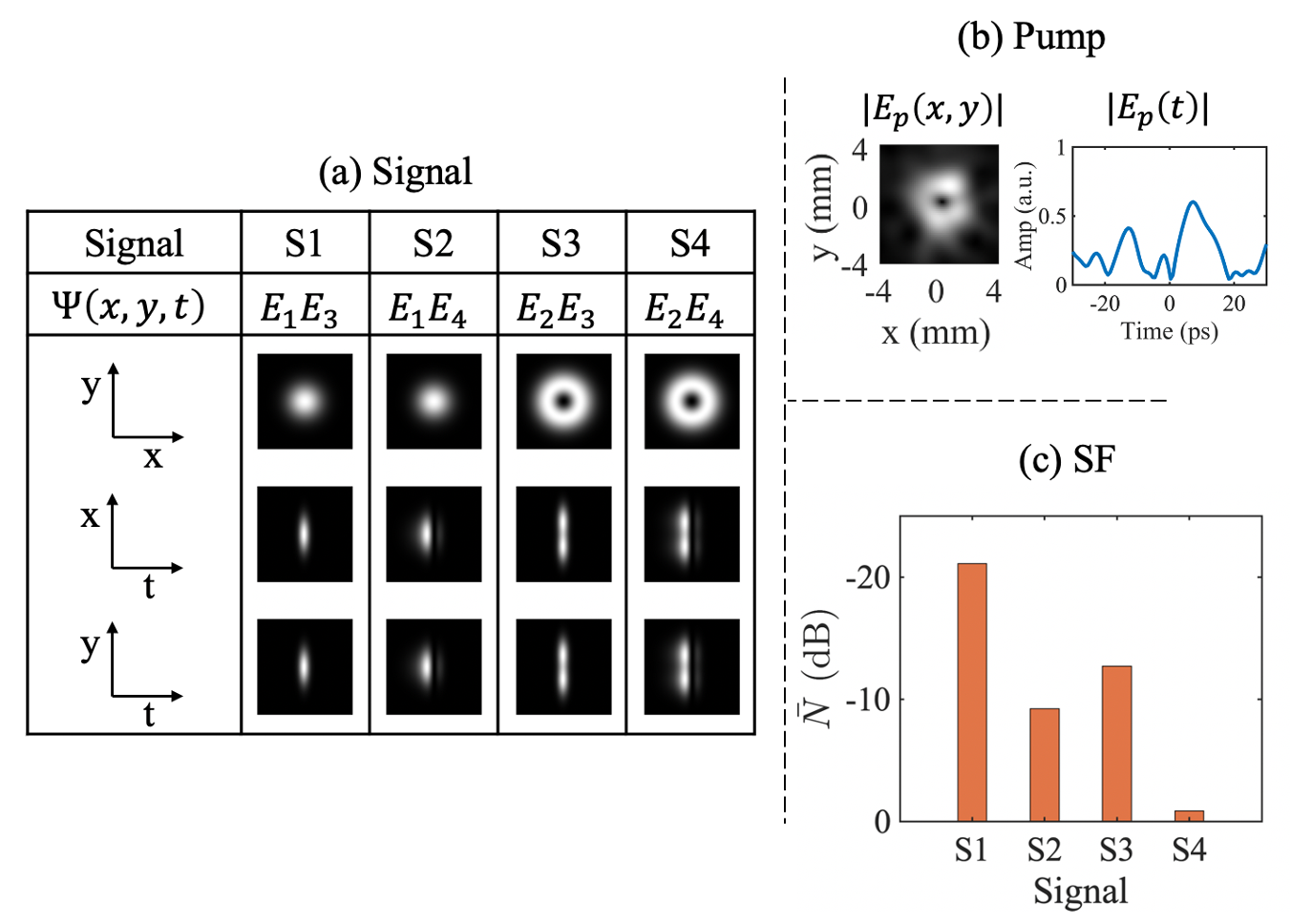}
    \caption{An example of selection among different spatio-temporal modes $S_k (k=1,2,3,4)$. (a) four different spatio-temporal modes $S_1$, $S_2$, $S_3$, and $S_4$, where the spatial and temporal components are shown in Fig. \ref{fig:Simulatedtimespatial}(a). (b) The optimized pump mode in the spatial and temporal domain. (c) Normalized SF photon for different signal modes with optimized pump mode. }
    \label{fig:Sim}
\end{figure}

\begin{figure*}[htbp]
\centering
\includegraphics[width=0.95\linewidth]{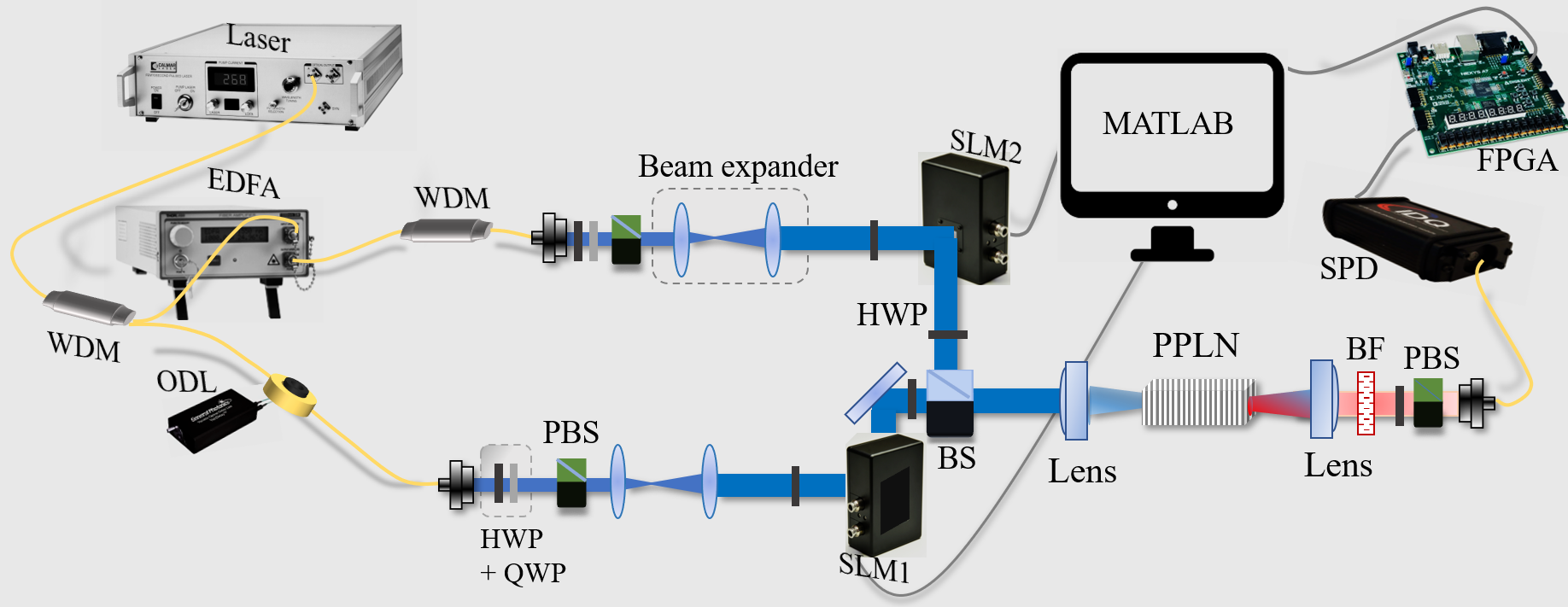}
\caption{ Experimental setup for high-dimensional mode selective frequency up-conversion process. ODL: programmable optical delay line, SLM: Spatial Light Modulator, BS: Beamsplitter, WDM: Wavelength-division multiplexing, PPLN: Magnesium-doped Periodic Poled Lithium Niobate crystal, BF: Shortpass bandpass filters and PM: power meter.}
\label{fig:setup}
\end{figure*}

Figure \ref{fig:Sim} shows an example in the simulation to select a signal mode ($S_4$) with respect to other overlapping signal modes ($S_1, S_2$, and $S_3$). For simplicity, we used a random-walk optimization method to optimize the pump field in spatio-temporal domain  \cite{Santosh19,QPMS2017}. We consider four signal modes $S_k (k=1,2,3,4)$, which are composed of two spatial ($E_1$, $E_2$) and two temporal ($E_3$, $E_4$) electric field amplitudes, as shown in Fig.~\ref{fig:Simulatedtimespatial}(a). One optimized pump, shown in Fig.~\ref{fig:Sim}(b), is designed to select $S_4$ and suppress the others simultaneously. With this optimized pump, the normalized up-converted SF output for different signals are shown in Fig.~\ref{fig:Sim}(c). The selectivity of the signal $S_4$ against $S_1$, $S_2$, and $S_3$ are 20.2 dB, 8.4 dB, and 11.8 dB, respectively. To further improve the selectivity, we can use a particle-swarm optimization algorithm in the spatial and temporal domain \cite{zhang_mode_2019}.


\section{Experimental Setup}
Figure \ref{fig:setup} outlines our present experimental setup. A mode-lock laser is used to generate the pulse train with pulse width $\sim$0.3 ps at 50 MHz repetition rate. We use two inline narrow-band wavelength division multiplexers (WDMs) with bandwidth 100 GHz to select two wavelengths, one at 1545 nm as a pump and another at 1558 nm as a signal. A programmable optical delay line (ODL) in the signal arm is used to scan the temporal shape of the optical pulses. The pump optical pulses are amplified using an Erbium-doped fiber amplifier (Thorlabs, EDFA100S) to obtain maximum average power $\sim$ 10 mW incident on the nonlinear crystal. On the other hand, the signal optical pulses are attenuated with neutral density filters to produce ultra-low optical signal. In free space, both arms select the horizontal polarization for the pump and signal. The transverse FWHM of the pump and signal are 2.8$\pm 0.05$ mm and 2.6$\pm 0.05$ mm, respectively. Then, these two beams are incident on separate spatial-light modulators (Santec SLM-100, 1440 $\times$ 1050 pixels, pixel pitch 10.4 $\times$ 10.4 $\mu$m) \cite{ZhangHE:20}. The pump and signal SLMs are used to upload desired (for example: HG or LG) phase masks. A beam splitter is used to combines the two pulse trains which are focused (focus length $F = $200 mm) into a temperature-stabilized second-order nonlinear crystal with a poling period of 19.36 $\mu$m (5 mol.\% MgO doped PPLN, 10 mm length, 3 mm width, and 1 mm height) for SF generation process. The normalized conversion efficiency of the nonlinear crystal is 1\%. Half waveplates on both arm, before the nonlinear crystal, are used to ensure the vertically polarized optical field parallel to the crystal's optical axis. The output pulses are then filtered with a short-pass filter to remove any residual fundamental light and bandpass filters to provide a total $>$180 dB extinction in rejecting the second harmonic of the pump \cite{QPMS2017,Rehain2020}. The visible SF photons pass through a half waveplate and polarizing beam splitter. The output SF photons are then coupled into a single-mode fiber (SMF-28) using a fiber collimator consisting of aspheric lens (Thorlabs A375TM-B). A silicon avalanche photodiode (Si-APD, ID100) is used to detect the photons which has 15\% quantum efficiency. The SF photons are then collected by the field programmable gate array (FPGA, Zedboard ZYNQ-7000) and sent to a computer for post processing. A MATLAB program is used to optimize the pump mode using adaptive feedback technique based on a random-walk optimization \cite{Santosh19,zhang_mode_2019}.

\section{Results}

\begin{figure}[hbt!]
    \centering
    \includegraphics[width=\linewidth]{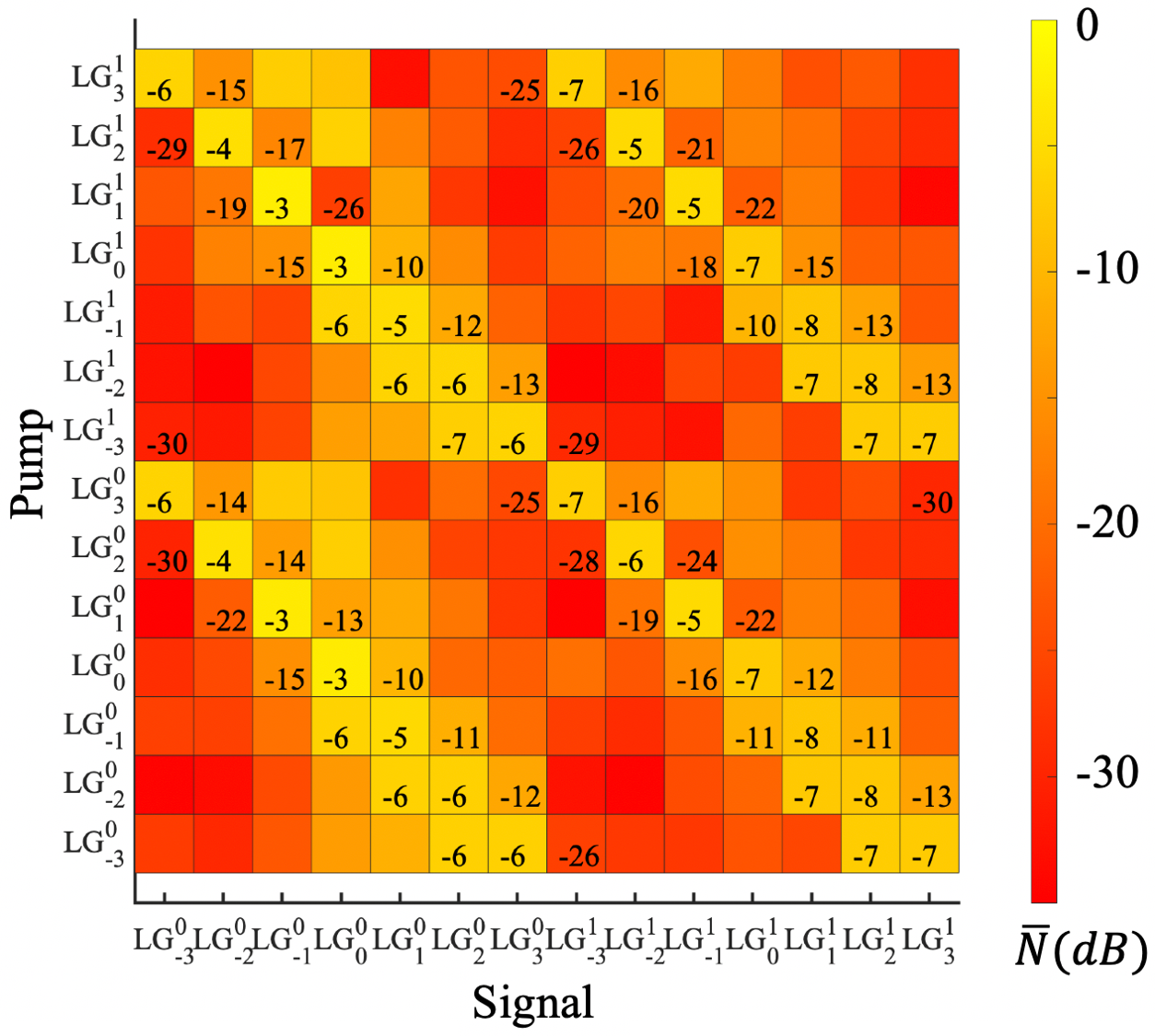}
    \caption{Measured SF output matrix with higher-order LG modes for the signal and the pump. The values in the matrix are representing the normalized SF output $\bar{N}$ dB (using Eq.\ref{eq:norm}).}
    \label{fig:LGmodes}
\end{figure}

Before presenting the results for mode-selective QFC with optimized pump, we first show the preparation of the quantum states in the spatial HD Hilbert space. Figure \ref{fig:LGmodes} shows the experimentally measured matrix for higher-order signal and pump modes. In this case, both signal and pump modes have the radial index $p=0,1$ and azimuthal index $l=-3$ to 3. The values in the color bar and matrix represent the normalized SF photon($\bar{N}$), which is normalized with the sum of the elements in each row (i.e. with same pump) according to Eq.\ref{eq:norm}. It is clear from this figure that the angular momentum with opposite sign for the signal and pump modes have higher extinction compare to other modes. We can also observe that the nearest neighbour spatial modes are difficult to distinguish \cite{Santosh19,Otte:20}. 

\begin{figure}[hbt!]
    \centering
    \includegraphics[height=5cm]{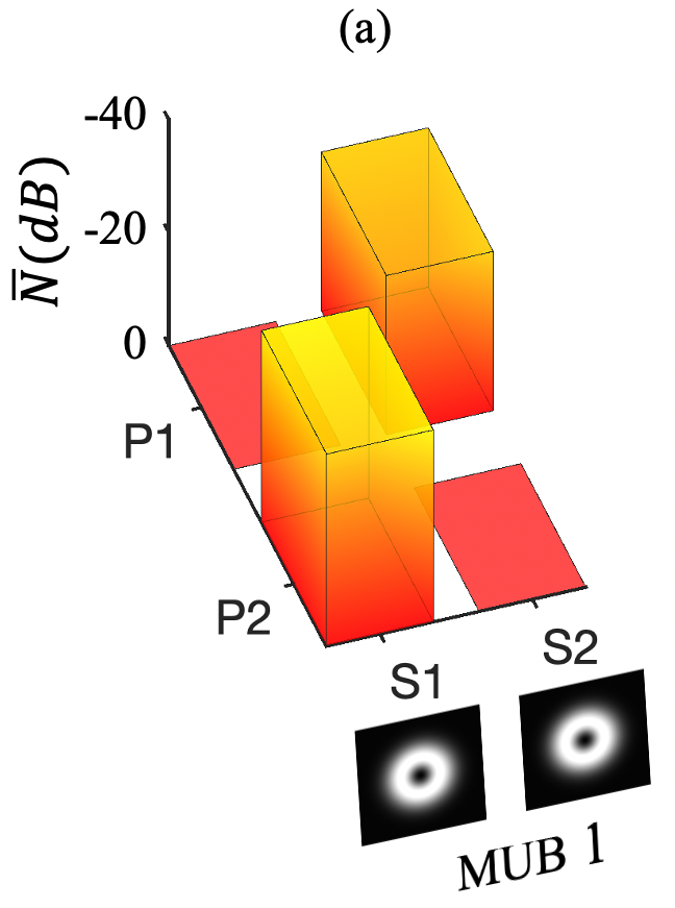}
    \includegraphics[height=5.1cm]{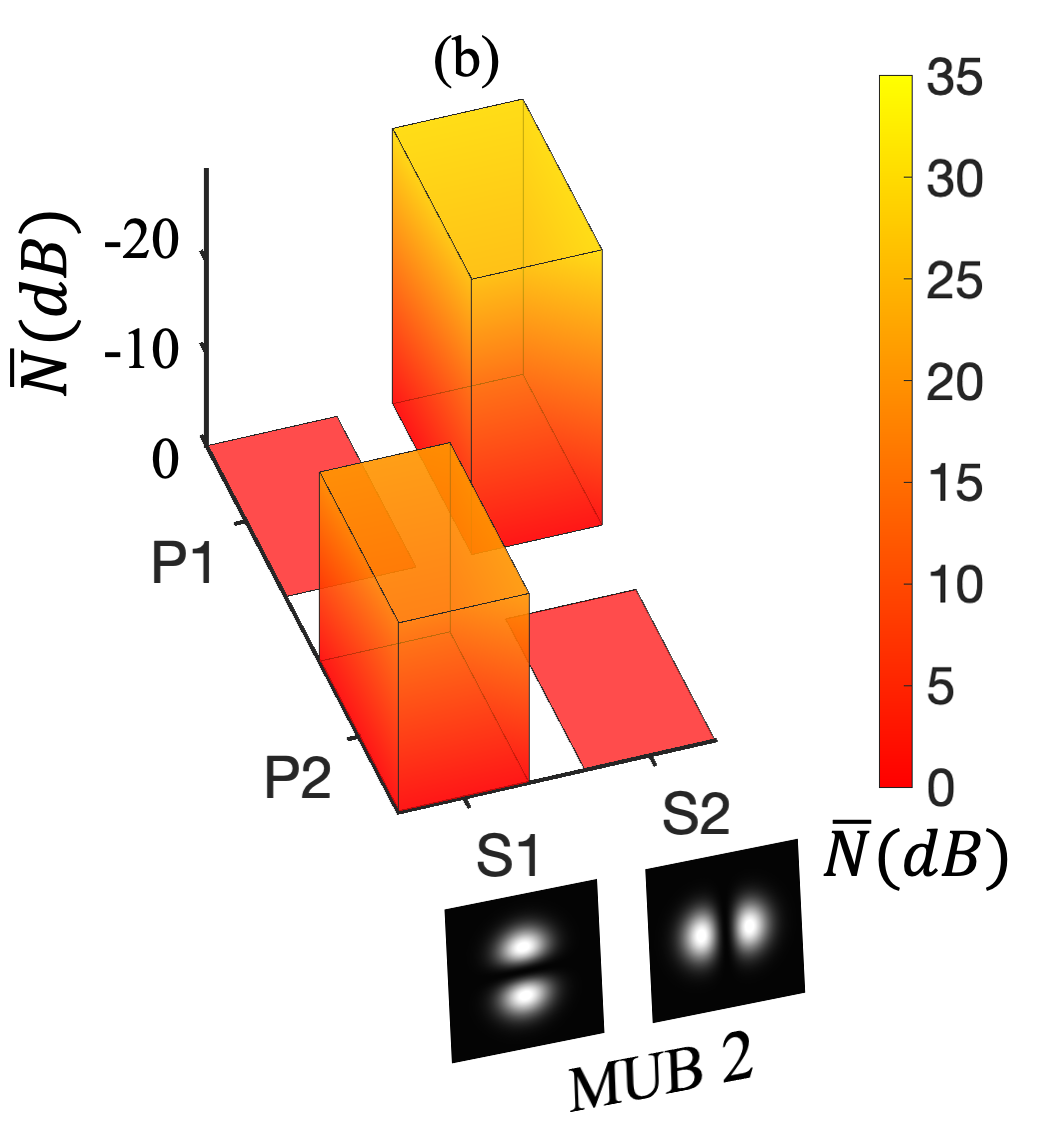}
    \caption{Qubit tomography for MUB of signal modes: (a) $\{|LG^0_{+1}\rangle, |LG^0_{-1}\rangle \}$ and (b) $\{(|LG^0_{+1}\rangle+|LG^0_{-1}\rangle)/\sqrt{2}, (|LG^0_{+1}\rangle-|LG^0_{-1}\rangle)/\sqrt{2} \}$. Normalization is done by summation of each row. $P_1$ and $P_2$ are the optimized pump modes to selectively upconvert signal $S_1$ and $S_2$, respectively.}
    \label{fig:Qubit}
\end{figure}

\begin{figure}[!htb]
    \centering
    \includegraphics[width=0.75\linewidth]{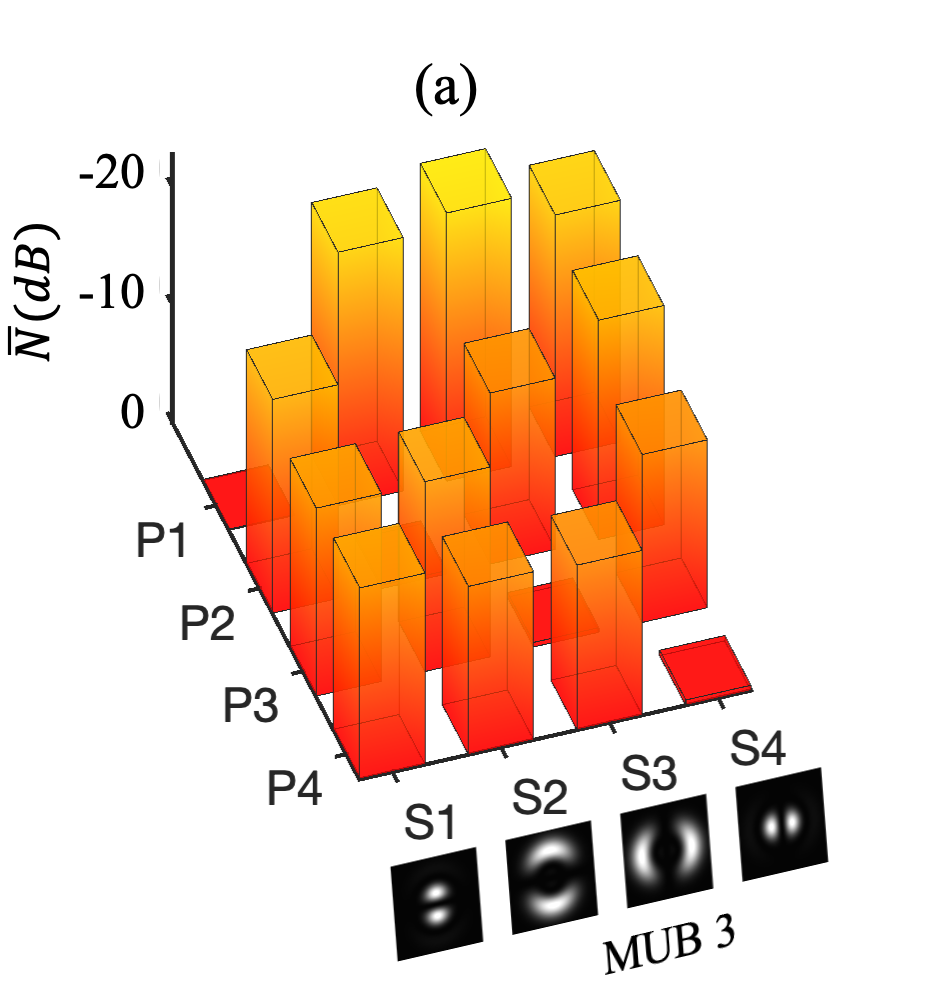}
    \includegraphics[width=0.8\linewidth]{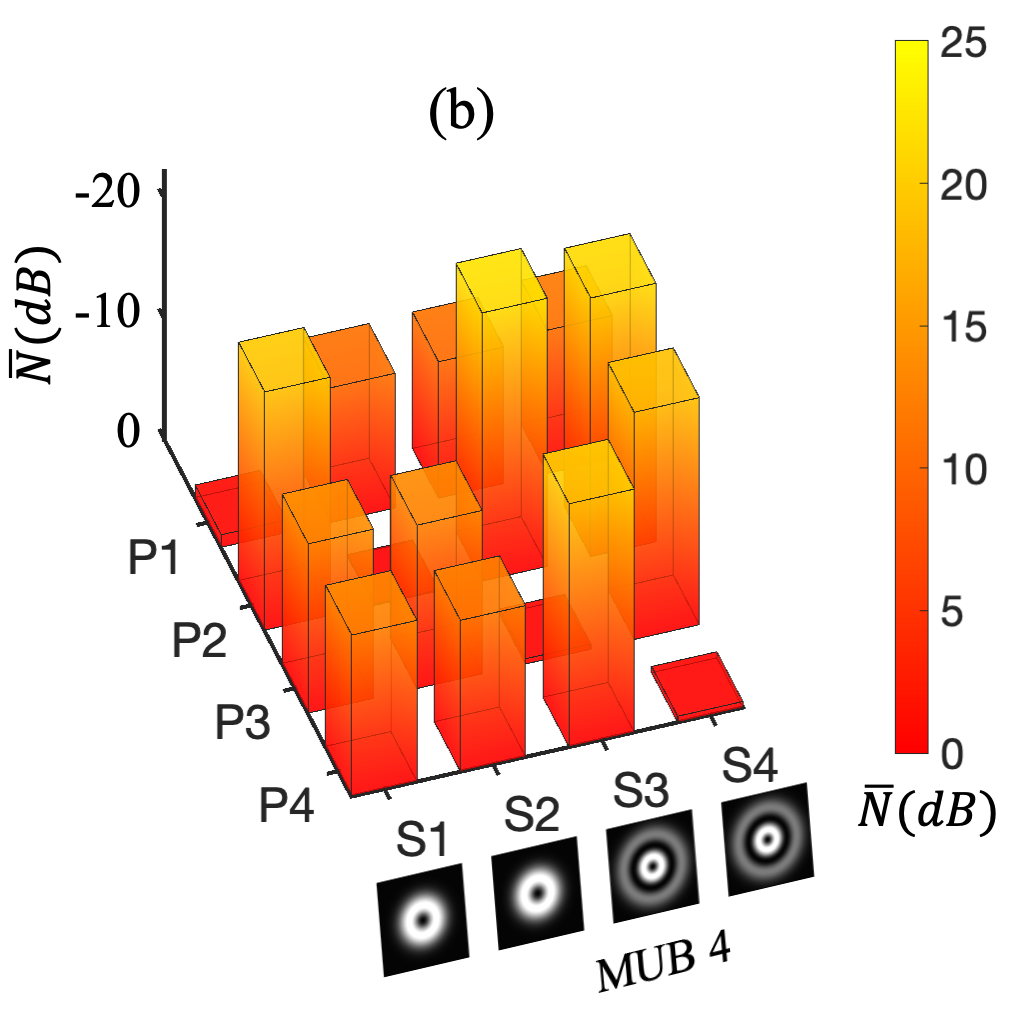}
    \caption{Qudit tomography for MUB  of signal modes: (a) $\{(|LG^0_{+1}\rangle+ |LG^0_{-1}\rangle + |LG^1_{+1}\rangle+ |LG^1_{-1}\rangle)/2, (|LG^0_{+1}\rangle+ |LG^0_{-1}\rangle - |LG^1_{+1}\rangle- |LG^1_{-1}\rangle)/2, (|LG^0_{+1}\rangle- |LG^0_{-1}\rangle - |LG^1_{+1}\rangle+ |LG^1_{-1}\rangle)/2, (|LG^0_{+1}\rangle- |LG^0_{-1}\rangle + |LG^1_{+1}\rangle- |LG^1_{-1}\rangle)/2 \}$ and (b) $\{|LG^0_{+1}\rangle, |LG^0_{-1}\rangle, |LG^1_{+1}\rangle, |LG^1_{-1}\rangle \}$. $P_i$ are the optimized pump modes to selectively upconvert signal $S_i$, where $i=1$ to 4.}
    \label{fig:Qudit}
\end{figure}

After that, optimized pump modes are prepared to selectively upconvert the MUB of qubit and qudit states. The optimized pump modes are in the superposition of the 18 higher-order LG modes ($l\in[-5:5],p\in[0,1]$). To show the selectivity for each optimized pump, the up-converted photons are collected and normalized. As an example, in Fig.~\ref{fig:Qubit}, we use two set of MUB for signal qubit states (a) \{$|LG^0_{+1}\rangle, |LG^0_{-1}\rangle$\} and (b) \{$(|LG^0_{+1}\rangle+|LG^0_{-1}\rangle)/\sqrt{2}, (|LG^0_{+1}\rangle-|LG^0_{-1}\rangle)/\sqrt{2}$\}. The qubit tomography shows that the optimized pump modes can selectively pickup the desired modes. It achieves the selectively of $\sim$ 28 dB on average. Similarly, Fig.~\ref{fig:Qudit} shows the superposition of 4 LG modes, i.e. qudit, to prepare 2 MUB consisting of 4 signal modes. Fig.~\ref{fig:Qudit} (a) shows the first MUB, we use signals \{$(|LG^0_{+1}\rangle+ |LG^0_{-1}\rangle + |LG^1_{+1}\rangle+ |LG^1_{-1}\rangle)/2, (|LG^0_{+1}\rangle+ |LG^0_{-1}\rangle - |LG^1_{+1}\rangle- |LG^1_{-1}\rangle)/2, (|LG^0_{+1}\rangle- |LG^0_{-1}\rangle - |LG^1_{+1}\rangle+ |LG^1_{-1}\rangle)/2, (|LG^0_{+1}\rangle- |LG^0_{-1}\rangle + |LG^1_{+1}\rangle- |LG^1_{-1}\rangle)/2$\} to selectively upconverts each with different optimized pump modes. The selectivity $\eta$ gives $\sim$17 dB on average extinction. In Fig.~\ref{fig:Qudit} (b), we show the second MUB, which consists 4 signals $|LG^0_{+1}\rangle, |LG^0_{-1}\rangle, |LG^1_{+1}\rangle$, and $|LG^1_{-1}\rangle$. We selectively upconvert these four signal with different optimized pump mode. The value of selectivity $\eta$ gives $\sim$16 dB extinction. 

\begin{figure}[ht]
    \centering
    \includegraphics[width=0.9\linewidth]{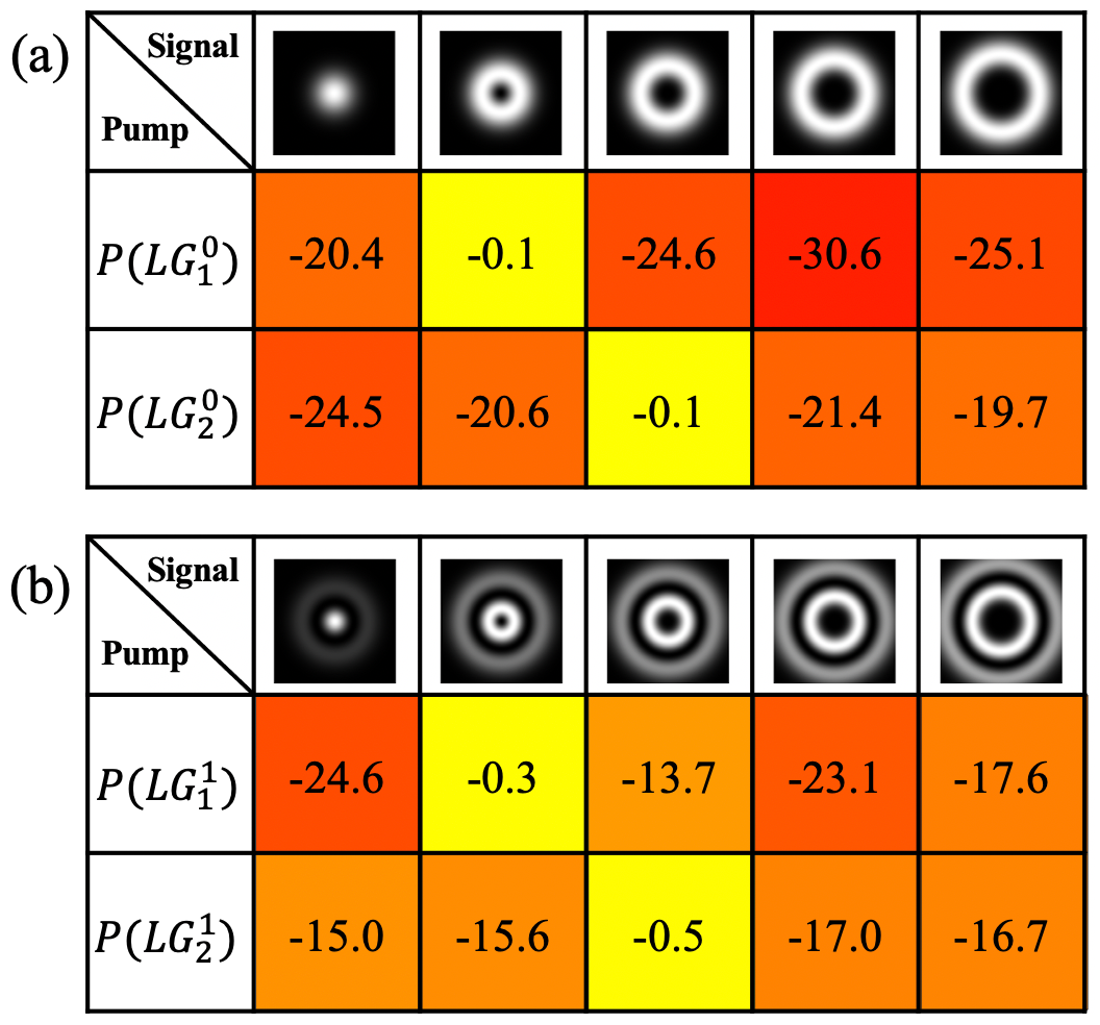}
    \caption{Spatial mode-selective upconversion of the signal modes with optimized superposition of the pump LG modes. The values in the table are the normalized SF output $\bar{N}$ dB, which are evaluated from Eq.\ref{eq:norm}. Five signal modes from left to right are (a) \{$LG_0^0$, $LG_1^0$, $LG_2^0$, $LG_3^0$, $LG_4^0$\}, and (b) \{$LG_0^1$, $LG_1^1$, $LG_2^1$, $LG_3^1$, $LG_4^1$\}, respectively.}
    \label{fig:Opt}
\end{figure}

Figure \ref{fig:Opt} demonstrates the selectivity among five higher-order LG modes (a) {$LG_0^0$, $LG_1^0$, $LG_2^0$, $LG_3^0$, $LG_4^0$} and (b) {$LG_0^1$, $LG_1^1$, $LG_2^1$, $LG_3^1$, $LG_4^1$} with two different optimized pumps, respectively. (a) and (b) show the selectivity of signal modes \{$LG_1^0$ \& $LG_2^0$ \}, and \{$LG_1^1$ \& $LG_2^1$\} with simultaneously suppressing other overlapping modes. Here also the selectivity for each optimized pump is determined by the collected and normalized SF photons, as defined in Eq. \ref{eq:selectivity}. We can see that the selectivity for 5 signal modes in Fig. \ref{fig:Opt}(a) and (b) are in between 20 dB to 30 dB and 15 dB to 25 dB, respectively.  





\begin{figure}[ht]
    \centering
    \includegraphics[width=0.9\linewidth]{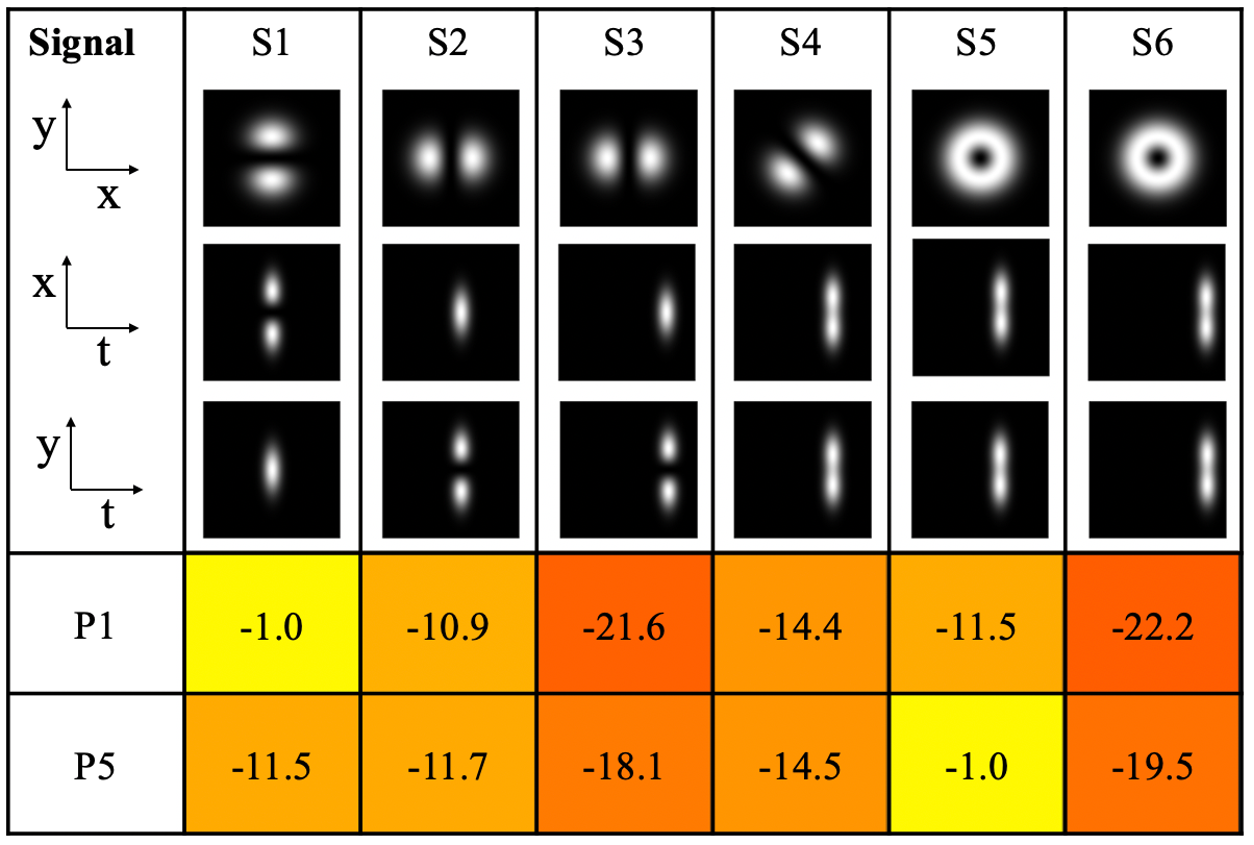}
    \caption{Spatio-temporal selectivity of the signal modes with optimized pump modes. The values in the table are the normalized SF output $\bar{N}$ dB. The spatial modes of these six signals are $S_1$: $(LG^0_{-1}+LG^0_{1})/\sqrt 2$, $S_2$: $(LG^0_{-1}-LG^0_{1})/\sqrt 2$, $S_3$: $(LG^0_{-1}-LG^0_{1})/\sqrt 2$, $S_4$: $(LG^0_{-1}-i LG^0_{1})/\sqrt 2$, $S_5$: $LG^0_{-1}$, and $S_6$: $LG^0_{1}$. The corresponding temporal delay, $t_0$, of the signals are $0$ ps, $5$ ps, $20$ ps, $15$ ps, $10$ ps, and $25$ ps, respectively.}
    \label{fig:Time-spatial}
\end{figure}

To show the performance of mode selective QFC in the spatial and temporal domain, we introduce the time-bin manipulation by changing the delay of the signal pulses with respect to the pump pulses. The upconverted photons are detected as a function of the temporal delay scanned by the programmable ODL between the synchronous signal and pump pulses. In Fig.~\ref{fig:Time-spatial}, we define signal modes in three dimensions, $x, y$, and $t$ and consider six signal modes. The spatial profile of these six signals are $S_1$: $(LG^0_{-1}+LG^0_{1})/\sqrt 2$, $S_2$: $(LG^0_{-1}-LG^0_{1})/\sqrt 2$, $S_3$: $(LG^0_{-1}-LG^0_{1})/\sqrt 2$, $S_4$: $(LG^0_{-1}-i LG^0_{1})/\sqrt 2$, $S_5$: $LG^0_{-1}$, and $S_6$: $LG^0_{1}$, and corresponding temporal delays are $t_0 = 0$ ps, $5$ ps, $20$ ps, $15$ ps, $10$ ps, and $25$ ps, respectively. The optimized pump $P_1$ ($P_5$) is designed to selectively up-convert $S_1$ ($S_5$) and suppress the others. The selectivities for $P_1$ and $P_5$ are 15 dB and 14 dB on average, respectively. These selectivities can further improve with other efficient adaptive optimization methods \cite{QPMS2017,PhysRevA_adaptive_Bayesian,Ansari:18}. 


\section{Discussion} 

In the current setup, the spatial mode shaping of the pump is provided by the SLM and the temporal shaping is realized by the time delay. In future, optical arbitrary waveform generator based on spectral line-by-line pulse shaping can be utilized to arbitrarily manipulate the amplitude and phase profiles of the time-frequency modes \cite{QPMS2017}. This can improve the selectivity of the signal modes by preparing the pump pulses in the optimized amplitude and phase profiles in time. Merging the selectivity in spatial and time-frequency modes will realize the full advantage of spatio-temporal mode selective QFC. We envision that spatio-temporal mode selectivity will be effective against channel noise in both measurable spatial and temporal dimensionality thus enables robust and realistic entanglement distribution with high-dimensional encoding \cite{PhysRevX.9.041042} with minimum trade-off. Note that, the selectivity and conversion efficiency of the QFC can be improved while suppressing it's intrinsic noise level by using longer nonlinear crystal with well defined phase matching profile \cite{manurkar_multidimensional_2016}. The signal-to-noise ratio in our current nonlinear crystal is characterized in Appendix B. 

Thanks to excellent selectivity among traverse spatial modes, our technique can be used for measurement of the traverse spatial mode joint probability distribution of a bi-photon state produced in spontaneous parametric down-conversion \cite{PhysRevA.69.023811,Ibarra-Borja:19} by performing mode selective measurement on the photon pair separately. Subsequently, this will allow one to directly quantify and certifying high-dimensional entanglement in traverse spatial mode.    

We can also incorporate polarization degree of freedom to extend the mode selective QFC \cite{Kaiser:19,PhysRevA.101.012339}. The preliminary measurements in the polarization degree of freedom, as shown in Appendix C, would extend the dimensionality of our mode-selective QFC.      


\section{Conclusion}
We have demonstrated a high-dimensional mode-selective QFC in spatio-temporal DOF. Our experimental results showed that a high-dimensional QFC can be realized with high-extinction (up to 30 dB) by selectively upconverting the signal modes using optimized pump. This technique is potentially useful for multi-fold quantum applications \cite{RMP_modes, Hu2020} such as long-distance quantum communication, quantum key distribution (QKD), quantum state tomography and so on. In future, it could be extended to other DOF to increase the information capacity of the quantum signals. 

\section*{Acknowledgement}
This research was supported in part by National Science Foundation (Grant No. 1842680) and in part by the Earth Science Technology Office, NASA, through the Instrument Incubator Program. 

\beginsupplement
\section*{Appendix}

\subsection{Spatial modes}
The Laguerre-Gaussian (LG) modes can be written in the cylindrical coordinates as
\begin{multline}
LG_l^{p}(r,\phi,z)=\frac{C_{lp}}{w}\left(\frac{r\sqrt{2}}{w}\right)^{|l|}L^{p}_{|l|}\left(\frac{2r^{2}}{w^{2}}\right)\exp(-il\phi)\\ \times \exp\left(i\zeta-ikz-\frac{ikr^{2}}{2R}-\frac{r^{2}}{w^{2}}\right),
\end{multline}

where $\phi=\arctan(y/x)$ is the azimuthal coordinate, ${\displaystyle C_{lp}}=\sqrt{\frac{2p!}{\pi(p+|l|)!}}$ is a normalization constant, $\{\displaystyle L^{p}_{\vert l \vert}\}$ are the generalized Laguerre polynomials with the azimuthal mode index $l$ and the radial index $p$. The LG modes can be expressed as the superposition of Hermite-Gaussian (HG) modes and vice-versa \cite{Wang_2016}. The $LG_l^p$ modes with the angular mode number $l = n-m$ and radial mode number $p = min(m, n)$. These can be written as

\begin{eqnarray}
 LG_{\pm1}^{0}=\frac{1}{\sqrt{2}}\left(HG_{01}\pm i HG_{10}\right), \\
 HG_{01}(\theta)= \frac{1}{\sqrt{2}}\left(LG_{+1}^{0} e^{-i\theta}+LG_{-1}^{0} e^{i\theta}\right),
\end{eqnarray}

where $\theta$ is the angle of rotation of the coordinate frame from $(x,y)$ to $(x',y')$.

\begin{figure}[!htb]
    \centering
    \includegraphics[width=0.49\linewidth]{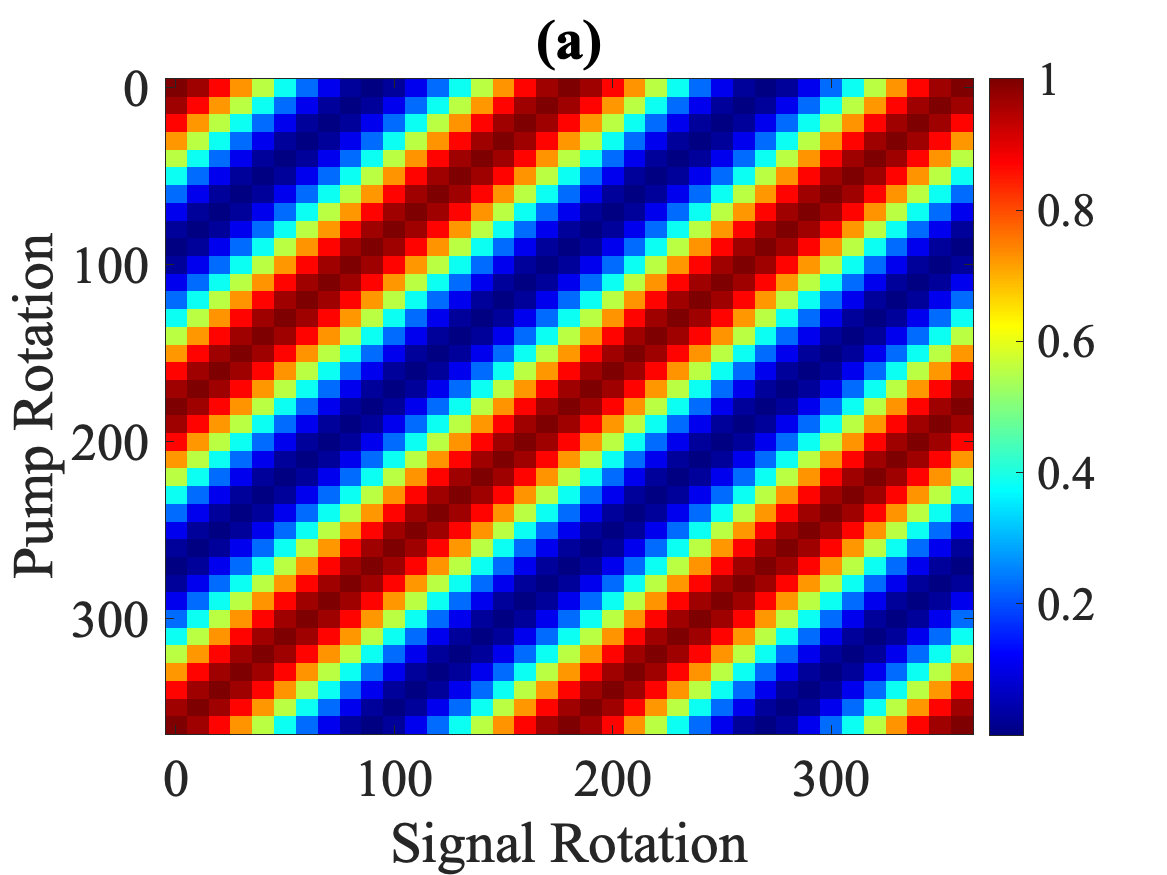}
    \includegraphics[width=0.49\linewidth]{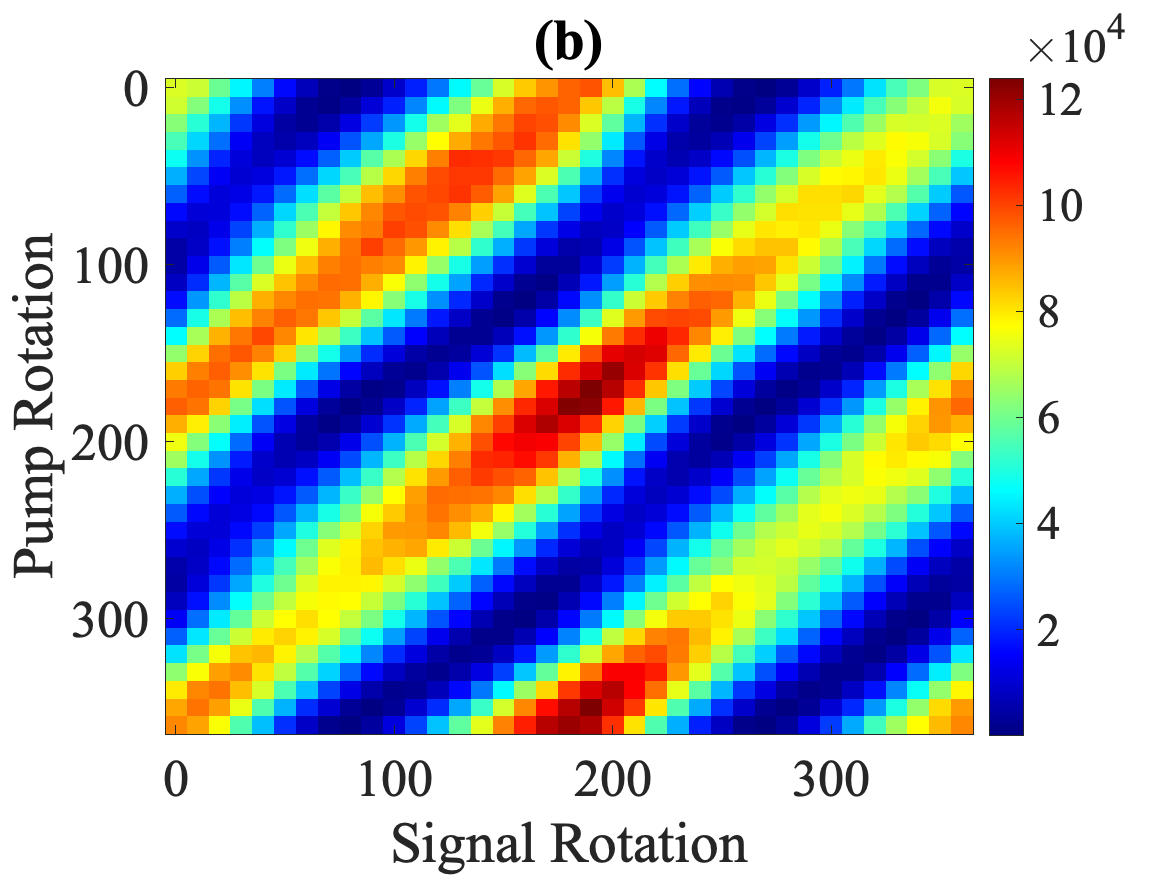}
    \caption{ SF photon counts are plotted by rotating the Hermite-Gaussian ($HG_{10}$) phase mask of the pump and the signal in (a) simulation and (b) experiment. The simulated result is normalized using the maximum value in the plot. The color bar, shown in (b), represents the detected SF counts. }
    \label{fig:mask_rotation}
\end{figure}

The rotation of the signal phase mask with respect to pump can manipulate the generated SF photons. As an example, in Fig.~\ref{fig:mask_rotation}, we plots the simulated and experimentally measured SF photons by rotating the $HG_{10}$ pump and signal masks. It shows that the SF photons are minimum, when the pump and signal phase masks are orthogonal to each other. This flexible control of phase mask rotation can be utilize for the HD quantum applications \cite{Fickler640,Moreaueaaw2563}.

\begin{figure}[hbt!]
    \centering
    \includegraphics[width=\linewidth]{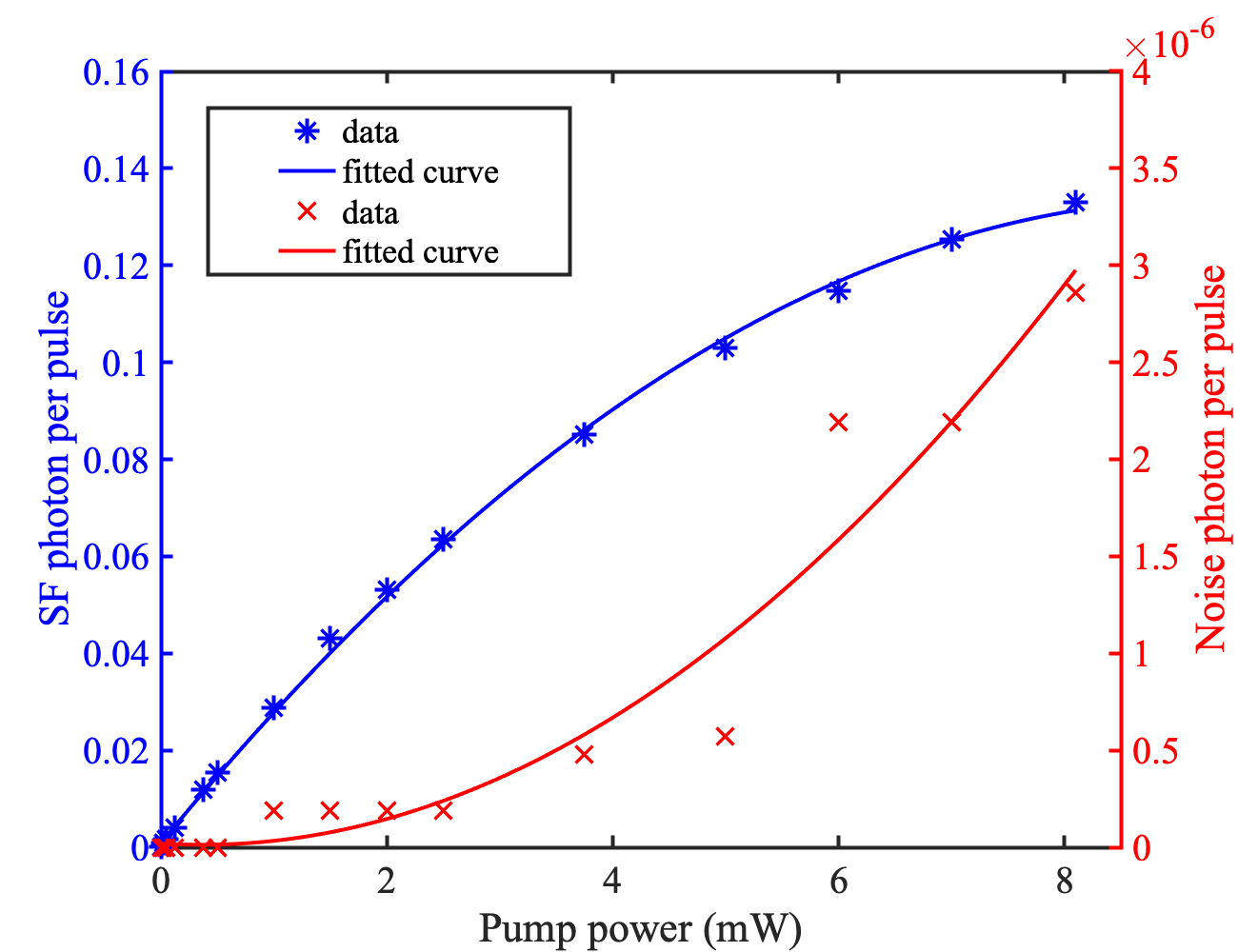}
    \caption{Experimentally measured SF photon counts (blue astric) and background noise photon counts (red cross) vs average pump power. Solid red and blue curves are the corresponding fits. 
}
    \label{fig_Raman}
\end{figure}

\subsection{Signal-to-Noise ratio}
We characterize the quantum noise generated in our nonlinear frequency conversion process in our experiment. We measured the upconverted SF photon per pulse and background noise photon per pulse vs average pump power as shown in Fig. \ref{fig_Raman}. The noise photon observed in Fig. \ref{fig_Raman} indicates that noise photon counts per pulse is $\sim$  $10^{-6}$, less than the noise level attained in LN waveguide \cite{2021arXiv210207044F}. The background noise photons could be generated by the pump field due to unwanted second harmonic and spontaneous parametric down conversion processes \cite{Sua2017,2021arXiv210207044F}. In this work, the background noise  $\sim$ 46 dB lower than the upconverted SF photons. 

\subsection{Polarization degree of freedom}
\begin{figure}[ht]
    \centering
    \includegraphics[width=\linewidth]{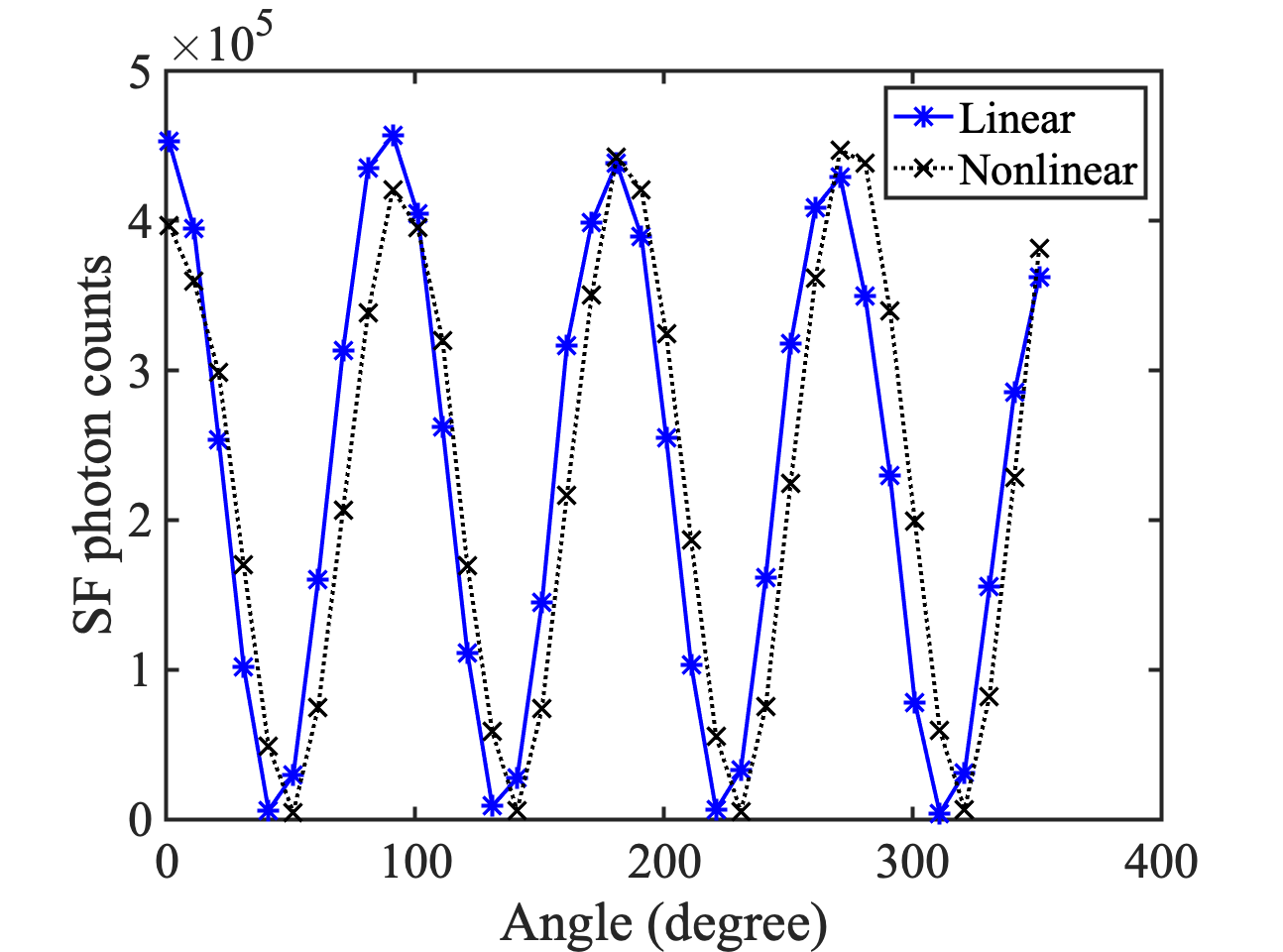}
    \caption{SF photon counts vs polarization rotation of the pump half-wave plate ($\theta_p$) with respect to the crystal. $\theta_{SF}$ is the polarization rotation of the half-wave plate with respect to the visible PBS.}
    \label{fig:polarization}
\end{figure}

We show the polarization dependence of the nonlinear crystal (black curve) and a visible PBS (blue curve) in Fig.~\ref{fig:polarization}. First measurement is performed by rotating the signal half-wave plate ($\theta_s$) in front of the nonlinear crystal and the second measurement (linear) is performed by rotating half-wave plate ($\theta_{SF}$) in front of visible PBS. In both cases, we record the photon counts using same visible APD detector. 
Visibility can be evaluated as  
\begin{equation}
V=\frac{C_{max}-C_{min}}{C_{max}+C_{min}},
\end{equation}
where $C_{max}$ and $C_{min}$ are the maximum and minimum number of SF photon counts. 

Our result in Fig.~\ref{fig:polarization} shows that the visibility of the photon counts in linear and nonlinear cases are 0.9832 and 0.9804, respectively. It shows that our system could be utilized for HD quantum states in polarization DOF \cite{Kaiser:19,Ren:21,PhysRevA.101.012339}. 

\bibliography{Ref}  

\end{document}